\documentclass[sigconf]{acmart}

\usepackage{threeparttable}
\usepackage{algorithm}
\usepackage{algorithmic}
\usepackage{booktabs}
\usepackage{enumitem}
\usepackage{multirow}
\usepackage{multicol}
\usepackage{array}
\usepackage{pifont}
\usepackage{listings}
\usepackage{amsmath}
\usepackage{color}
\usepackage{xcolor}

\usepackage{soul, color, xcolor,xspace}
\newcommand{\name}{UQABench\xspace}

\AtBeginDocument{%
  }

\setcopyright{acmlicensed}
\copyrightyear{2018}
\acmYear{2018}
\acmDOI{XXXXXXX.XXXXXXX}

\acmConference[Conference acronym 'XX]{Make sure to enter the correct
  conference title from your rights confirmation emai}{June 03--05,
  2018}{Woodstock, NY}
\acmISBN{978-1-4503-XXXX-X/18/06}

\begin{document}

\title{UQABench: Evaluating User Embedding for Prompting LLMs in Personalized Question Answering}



\author{Langming Liu$^{\dagger}$, Shilei Liu$^{\dagger}$, Yujin Yuan, Yizhen Zhang, Bencheng Yan, Zhiyuan Zeng, Zihao Wang, Jiaqi Liu, Di Wang, Wenbo Su, Wang Pengjie, Jian Xu, Bo Zheng}
\affiliation{%
  \institution{Taobao \& Tmall Group of Alibaba}
  \city{}
  \country{}}




  



\renewcommand{\shortauthors}{Langming Liu, et al.}

\begin{abstract}
Large language models (LLMs) achieve remarkable success in natural language processing (NLP). In practical scenarios like recommendations, as users increasingly seek personalized experiences, it becomes crucial to incorporate user interaction history into the context of LLMs to enhance personalization. 
However, from a practical utility perspective, user interactions' extensive length and noise present challenges when used directly as text prompts. 
A promising solution is to compress and distill interactions into compact embeddings, serving as soft prompts to assist LLMs in generating personalized responses. Although this approach brings efficiency, a critical concern emerges: Can user embeddings adequately capture valuable information and prompt LLMs?
To address this concern, we propose \name, a benchmark designed to evaluate the effectiveness of user embeddings in prompting LLMs for personalization. We establish a fair and standardized evaluation process, encompassing pre-training, fine-tuning, and evaluation stages. To thoroughly evaluate user embeddings, we design three dimensions of tasks: sequence understanding, action prediction, and interest perception. These evaluation tasks cover the industry's demands in traditional recommendation tasks, such as improving prediction accuracy, and its aspirations for LLM-based methods, such as accurately understanding user interests and enhancing the user experience.
We conduct extensive experiments on various state-of-the-art methods for modeling user embeddings. Additionally, we reveal the scaling laws of leveraging user embeddings to prompt LLMs.  
The benchmark is available online at~\url{https://github.com/OpenStellarTeam/UQABench}.

\end{abstract}



\maketitle

\sloppy

\section{Introduction}
The advent of large language models (LLMs) has revolutionized the field of natural language processing (NLP), showcasing remarkable performance across various tasks, such as machine translation, sentiment analysis, semantic understanding, and multi-turn dialogues~\cite{OpenAI2023GPT4,touvron2023llama,yang2024qwen2,liu2024deepseek}. 
The powerful capabilities of LLMs have the potential to extend to numerous personalized scenarios, including platform search, recommendation, and online advertising, which are highly related to user experience~\cite{guu2020retrieval,bao2023tallrec,muhamed2021ctr}. Specifically, the information cocoons caused by collaborative filtering (CF)-based methods are always a critical challenge that jeopardizes user experience in personalization tasks~\cite{pariser2011filter,nguyen2014exploring}.
With rich worldwide knowledge and diversity, LLMs provide a chance to break the information cocoons of users and bring sufficient personalization~\cite{xu2024prompting,zhao2023recommender,zhang2023recommendation,wu2024survey}. One promising direction LLMs can take to provide personalization is incorporating user history interactions as contextual information~\cite{liu2023chatgpt,petrov2023generative,kang2023llms,geng2022recommendation,lyu2023llm}. Specifically, the user interaction data are collected from the behavioral interactions (e.g., click, favorite, add to cart, and purchase) between users and items in various scenarios, e.g., searching and recommendation. Therefore, user interaction data can provide abundant information regarding user interests and preferences, a valuable source for providing user-side context for prompting LLMs in personalization.

Despite the promise of enhanced personalization by integrating user interaction history into the context of LLMs, this approach faces significant challenges. User interaction data can be lengthy and noisy~\cite{wang2021denoising,tian2022learning,qin2021world}, making direct incorporation as text prompts impractical in industrial settings.
First, the massive number of tokens in interaction sequences (e.g., tens of thousands) slows down the inference speed of LLMs, leading to unacceptable response times and negatively impacting user experience. Technically, the computational complexity of the attention mechanism---the core of LLMs---is enormous, scaling quadratically with the number of tokens $N$~\cite{vaswani2017attention}. Moreover, $N$ may even exceed the context limit of LLMs.
Second, users' redundant, repeated, and unintentional interactions across various scenarios can mislead LLMs into understanding user interests.
Inspired by advancements in recommendations, such as GRU4Rec~\cite{hidasi2015session} and SASRec~\cite{kang2018self}, a promising alternative to address this issue is compressing these interactions into more compact user embeddings using encoder models~\cite{li2023prompt,ning2024user,li2023personalized}. These user embeddings act as soft prompts, providing ``efficient'' (i.e., fewer tokens, higher value) contextual information. By extracting high-value information and filtering out irrelevant interactions, this embedding-based approach simultaneously addresses the aforementioned challenges.


While employing user embeddings to prompt LLMs offers a promising and efficient approach to personalization, several significant concerns emerge regarding their effectiveness: 
\begin{itemize}[leftmargin=*]
\item Can user embeddings capture the necessary information from user interactions and effectively convey it to LLMs? 
\item When contextualized through user embeddings, can LLMs perform traditional recommendation tasks successfully?
\item Do user embeddings serve as strong prompts that guide LLMs to produce personalized responses aligned with user interests?
\end{itemize}
Unfortunately, previous evaluations primarily concentrate on the traditional recommendation performance~\cite{fang2020deep,hidasi2015session,kang2018self}, such as user-item similarity scores and recall/rank metrics.
This focus leaves a gap in comprehensively assessing how user embeddings address these concerns in the LLM era.
To bridge this gap, we introduce \textbf{\name}, a novel Chinese \textbf{Bench}mark designed to evaluate the quality of \textbf{U}ser embeddings in prompting LLMs for personalized \textbf{Q}uestion \textbf{A}nswering. The interactive, personalized Q\&A paradigm aligns with the LLMs' parameterized nature. 
Notably, our approach stands out for its standardization and comprehensiveness in evaluating user embeddings.

The standardized evaluation flow of \name includes three steps: 
(1) \textbf{Pre-training}: We utilize sufficient user interaction data to pre-train encoder models. 
(2) \textbf{Fine-tuning}: We align learned user embeddings with semantic space through fine-tuning, a step crucial for making LLMs capable of utilizing user embeddings for personalization.
(3) \textbf{Evaluating}: We evaluate user embeddings in prompting LLMs for task-specific personalized Q\&A.  
To provide a comprehensive assessment, we design three critical tasks that extend beyond conventional evaluations and address the previously mentioned concerns:
(1) \textbf{Sequence understanding} measures the qualities of user embeddings in prompting LLMs to understand direct features and match features, crucial for determining whether user embeddings can substitute massive feature engineering in the industry.
(2) \textbf{Action prediction} involves next-item and next-attribute predictions, reframing traditional recommendation tasks within a natural language paradigm.
(3) \textbf{Interest perception} focuses on modeling user interest, covering long- and short-term interests and interest trajectory, aligning closely with new industrial demands such as enhancing user experience. 
We evaluate various state-of-the-art user encoder models using \name, providing valuable insights for employing user embeddings in the LLM era. Furthermore, we explore the scalability of Transformer-based encoders in prompting LLMs for personalization.
Our work seeks not only to validate effective methodologies but also to lay the groundwork for future innovations in user personalization.

The main contributions of our work are three-fold:

\begin{itemize}[leftmargin=*]

\item We introduce \name, a novel benchmark that comprehensively evaluates the user embeddings in prompting LLMs for personalization. This benchmark uniquely leverages personalized interactive question answering, distinguishing itself from traditional benchmarks.


\item We construct a standardized assessing process to ensure evaluation reliability. 
Additionally, our proposed assessment tasks comprehensively address the effectiveness of user embeddings, encompassing industrial recommender system focuses from both old and new eras.

\item The data and code are open source, facilitating further research. Extensive experiments offer novel insights into learning user representations, providing guidelines for optimizing personalization and enhancing user experience in the LLM era.  

\end{itemize}




\begin{figure}[t]
    \centering
    \includegraphics[width=1.0\linewidth]{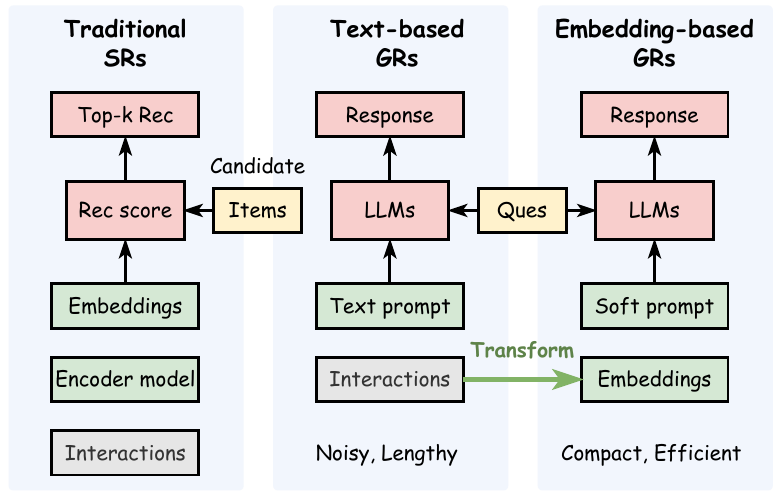}
    \caption{
    A brief comparison of SRs and GRs.
    }
    \label{fig:compare}
    \vspace{-2mm}
\end{figure}

\section{Preliminary}
We briefly introduce how both previous and current recommender systems utilize user interactions, focusing on two prevailing branches: sequential recommendations (SRs) and generative recommendations (GRs). The brief comparison is illustrated in Figure~\ref{fig:compare}.

\subsection{Sequential Recommendations (SRs)}
Given a set of users \(\mathcal{U}=\{u_1,u_2,\cdots,u_{\vert\mathcal{U}\vert}\}\) and a set of items \(\mathcal{V}=\{v_1,v_2,\cdots,v_{\vert\mathcal{V}\vert}\}\), consider that the interaction sequence of user \(u_i\) is denoted as \(s_i=[v_{1}^i, v_{2}^i, \cdots, v_{n_i}^i]\). For user \(u_i\), SRs take \(s_i\) as input and calculate recommendation scores (e.g., cosine similarity) for candidate items, then output the top-\(k\) items most likely to be interacted with in the subsequent time step.

\subsection{Generative Recommendations (GRs)}
GRs represent a shift towards utilizing LLMs to generate more personalized recommendation results, divided into two main branches:

\subsubsection{\textbf{Text-based GRs.}} 
Text-based GRs directly leverage the textual information of interactions. Interacted items' IDs and textual attributes, such as titles, categories, and descriptions, are chronologically placed in pre-designed prompts to serve as user context. Given questions like "What item will the user click next," the LLMs generate responses such as "item $j$."

\subsubsection{\textbf{Embedding-based GRs.}}
In contrast, embedding-based GRs transform lengthy, noisy interaction sequences into information-dense user embeddings. These user embeddings, aligned to the semantic space by an adapter, act as soft prompts for LLMs, personalizing them further. Compared to text-based GRs, this approach requires fewer tokens to convey user context, enhancing efficiency and scalability. However, a critical question remains: "Can this approach provide comparable personalized performance to text-based GRs?" This question forms one of the core research focuses of this paper.

\begin{figure*}[t]
    \centering
    \includegraphics[width=0.9\linewidth]{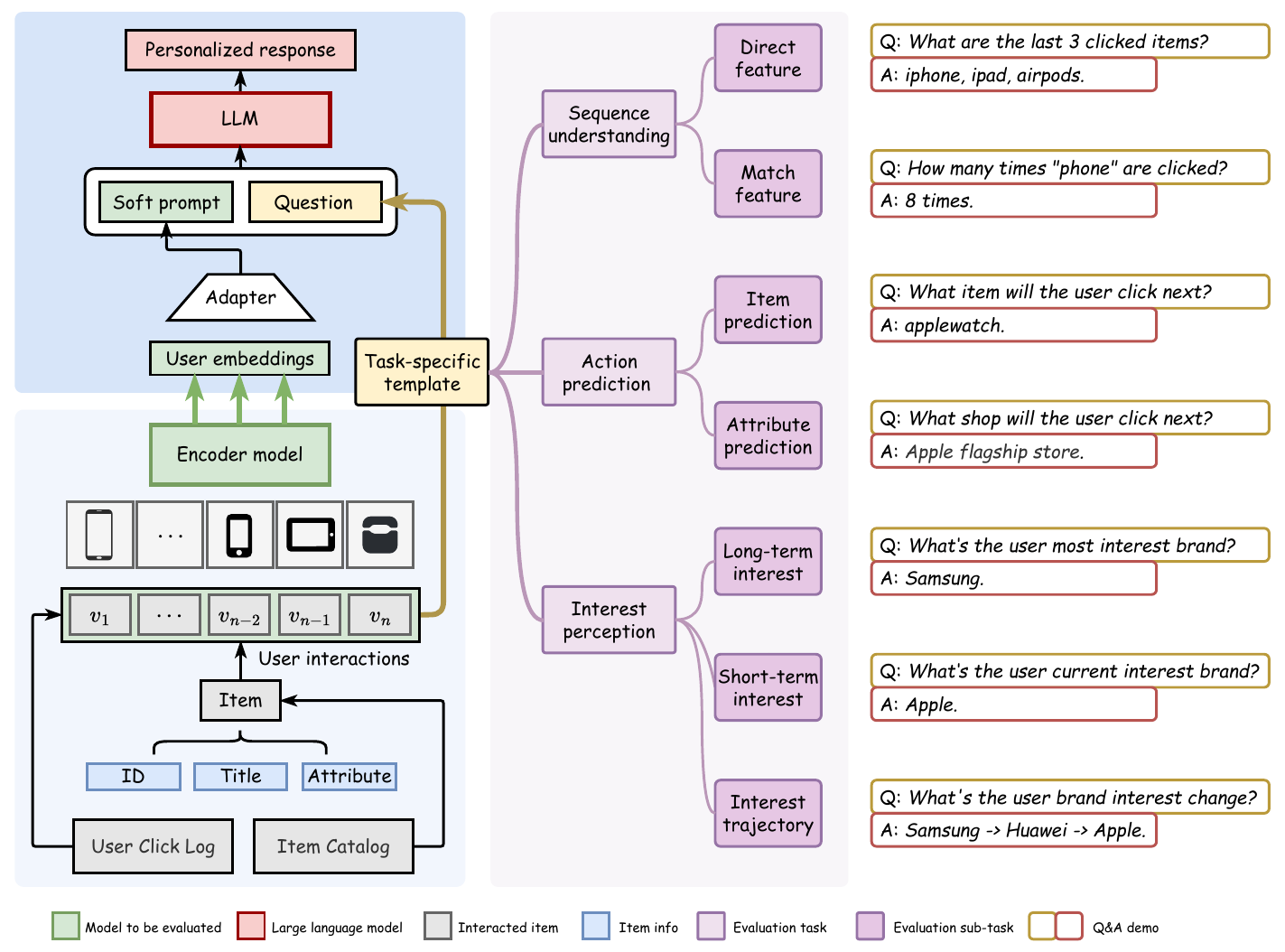}
    \caption{
    An overview of the standardized evaluation process of user embeddings on \name. The left-bottom part demonstrates the encoder model encodes the historical interactions into user embeddings, which requires pre-training. The left-top part shows that user embeddings act as the soft prompt for LLMs to generate personalized responses, which needs fine-tuning. The right side demonstrates the proposed tasks in the evaluating stage and the corresponding Q\&A demo.  
    }
    \label{fig:overview}
\end{figure*}

\section{\name}
The overview of \name is shown in Figure~\ref{fig:overview}. Based on the industrial user click log and item catalog, we construct the user interaction data as the evaluation source. The evaluation process comprises three critical steps: \textbf{pre-training}, \textbf{fine-tuning}, and \textbf{evaluating}. 
First, we pre-train the encoder models on user interactions to learn initial user embeddings.
Next, we align these user embeddings to the semantic space through fine-tuning, enabling them to act as soft prompts for LLMs. Finally, we assess the quality of the embeddings by evaluating the performance of LLMs prompted by these embeddings in personalized question answering tasks.

\subsection{Source Data}
To build a benchmark specialized in user personalization, we require an open source dataset that includes sufficient user interactions and item information. Unfortunately, two crucial reasons hinder us from directly leveraging existing datasets, such as MovieLens, Yelp, and Amazon Review Data.
First, such datasets mainly focus on specific fields (e.g., movies, merchants, books, and sports) and specific tasks (e.g., next-item and rating predictions), jeopardizing evaluation comprehensiveness.
Second, these datasets suffer from issues related to information richness, such as limited users and items, insufficient side information, and short interaction sequences, all of which reduce the reliability of the evaluation.
To address the issues above, we construct and release a brand-new user interaction dataset as the source data for \name. This dataset is collected and processed from \textbf{Taobao's User Click Log}. Statistically, it includes $184,520$ users, $994,447$ items, and $31,317,087$ interactions, along with extensive item-side information (e.g., item title, category name, brand name, shop name) sourced from \textbf{Taobao's Item Catalog}. 
To ensure user privacy and comply with regulations, we perform necessary operations, detailed in Section~\ref{sec:data}, on data without destroying the sequence patterns. Then, we launch the evaluation flow leveraging the well-prepared source data. 


\subsection{Stage 1: Pre-training}
The first step of evaluation on \name is the pre-training stage, which is essential for encoder models to learn to encode noisy and lengthy user interaction sequences into denser representations (i.e., user embeddings). In this stage, we focus on three critical factors that greatly affect the upper bound of performance: model inputs, training tasks, and training methods. To determine the optimal settings of these three critical parts, we conduct a corresponding ablation study in Section~\ref{sec:ablation_1}.

\subsubsection{\textbf{Model Inputs.}}
Model inputs are fundamental in pre-training encoder models. The item ID, serving as the unique identifier for each item (e.g., book, movie, music), remains the most critical feature in traditional recommendations. Additionally, side information such as title, category, brand, and shop provides crucial contextual information that enhances recommendation performance. The item side information includes ID-based information (e.g., category ID) and text-based information (e.g., category name). 
We utilize item ID as the necessary input and consider the ID-based and text-based side information as the optional inputs. 

\subsubsection{\textbf{Training Tasks.}}
Determining the appropriate training task is vital for the pre-training stage. Although different encoder models may use various tasks (e.g., sequential or bidirectional order, next prediction or cloze tasks, item or attribute prediction) in the original works, we standardize the training task to ensure fairness and improve efficiency. We adopt the next-item prediction (NIP)~\cite{hidasi2015session,kang2018self} as the uniform pre-training task for all encoder models. NIP is a widely used paradigm in sequential recommendations and has proven its effectiveness in learning user representations~\cite{fang2020deep}.

\subsubsection{\textbf{Training Methods.}}
Regarding encoder training methods, two common approaches in recommender systems are employed: \textbf{supervised learning} and \textbf{contrastive learning}. 
Supervised learning (SL), typically optimized via classification losses such as cross-entropy (CE), achieves strong performance when labeled data is abundant and reliable.
However, in real-world scenarios where explicit labels are scarce, and the item pool is vast (leading to data sparsity), SL struggles to generalize. In contrast, contrastive learning (CL) methods such as InfoNCE are preferred in industrial settings.
Contrastive learning operates under a self-supervised paradigm: it learns user and item representations by contrasting positive pairs (e.g., user-item interactions) against negative pairs (e.g., randomly sampled non-interactions).
This method achieves efficiency via in-batch negative sampling and robustness in cold-start/long-tail scenarios by leveraging intrinsic data structures.

\subsection{Stage 2: Fine-tuning}
The pre-trained encoders are capable of producing user embeddings, capturing user interests and sequential patterns.
Unlike serving collaborative filtering functions (e.g., calculating user-item similarity) in traditional recommendations, user embeddings will act as a soft prompt to activate the personalization of LLMs. However, the discrepancy between the dimensions of the user embeddings and semantic space obstructs direct use. 
To this end, the adapter is needed to transform user embeddings into word embeddings' dimensions. Additionally, compression methods are adopted to reduce the length of user embeddings, where we consider two approaches: \textbf{Mean Pooling} and \textbf{Q-Former}, specified in Section~\ref{sec:ablation_2}. In this way, the user embeddings are converted to several ``soft-prompt tokens.''
A fine-tuning stage of the framework is necessary for (1) transferring the information encoded in user embeddings to LLMs and (2) generating responses consistent with the answer paradigm to improve evaluation efficiency.
We freeze the pre-trained encoders to avoid the knowledge of LLMs overwhelming them and causing catastrophic forgetting of learned user sequential information.
We explore two fine-tuning strategies: full fine-tuning (LLM+adapter) and partial fine-tuning (only adapter) in Section~\ref{sec:ablation_2}.


\begin{table*}
\caption{\label{tab:task_intro}
Details and interpretations of evaluation tasks.}
    \centering
    \begin{threeparttable}
    \begin{tabular}{lll}
        \toprule
        \toprule
        \textbf{Task} & \textbf{Sub-task} & \textbf{Interpretation} \\
        \cmidrule(lr){1-3}
        {Sequence understanding} 
        & Direct feature (DF)       & Directly recovering the features in user interaction sequence. \cr
        & Match feature (MF)        & Counting the statistics of features matching the target feature.  \cr
        \cmidrule(lr){1-3}
        {Action prediction} 
        & Item prediction (IP)      & Predicting the next item the user will most likely interact with. \cr
        & Attribute prediction (AP) & Predicting the attribute of the next item the user will most likely interact with. \cr
        \cmidrule(lr){1-3}
        {Interest perception} 
        & Long-term interest (LI)   & Perceiving the user's long-term interest category/brand/store. \cr
        & Short-term interest (SI)  & Perceiving the user's current interest category/brand/store. \cr
        & Interest trajectory (IT)  & Detecting the interest changes of the user. \cr
        \bottomrule
        \bottomrule
    \end{tabular}
    \end{threeparttable}
    \vspace{-2mm}
\end{table*}

\subsection{Stage 3: Evaluating}
Next, we prompt the LLMs with aligned user embeddings to generate personalized responses and evaluate their quality. 
The critical challenge is scaling up the evaluation while guaranteeing objectivity simultaneously. 
To this end, we manually design some question templates as seeds according to the interaction data patterns and expertise. 
This way, massive question prompts (with their ground truths) can be automatically generated based on user historical interactions.
We propose three critical evaluation tasks to assess user embeddings from multi-dimensional perspectives: sequence understanding, action prediction, and interest perception, as illustrated in Figure~\ref{fig:overview}. 
As specified in Table~\ref{tab:task_intro}, we subdivide each task into two or three sub-tasks for a more fine-grained evaluation and design various task-specific templates for each sub-task. Some details of template designing are mentioned in Section~\ref{sec:prompt}.
We then introduce each evaluation task from a conceptual perspective.


\subsubsection{\textbf{Sequence Understanding.}}
Sequence understanding involves using LLMs to extract and restore historical user information from user embeddings.
The primary goal of this task is to assess how well user embeddings can serve as a bridge to convey necessary information from user interaction sequences to LLMs.
This task is also of significant interest to the industry since it is related to whether user embeddings in the LLM era can substitute for massive feature engineering on the user side.
Specifically, for a given user interaction sequence, we can query LLMs for  \textbf{direct features} like item IDs, titles, and other side information such as category, shop, and brand.
For example, we can construct the question prompt such that, "\textit{what are the last three items the user has clicked?}" or, "\textit{what is the brand of the last item the user has clicked?}"
In user modeling of industrial recommendation scenarios, the \textbf{match features} play a critical role in feature engineering that captures high-order user sequence patterns.   
To this end, we design questions focusing on match features, defined as the statistical attributes of the user sequence that match the target feature, like item, category, or brand.
For example, "\textit{given a category, tell me the times the user has clicked on this category in history}." Although the questions about match features are more complicated than direct features, it is a practical approach for exploring the upper bound of user embeddings in the sequence understanding task.


\subsubsection{\textbf{Action Prediction.}}
We are interested in how user embeddings can assist LLMs in completing traditional industrial tasks like top-$k$ recommendations and click-through rate (CTR) prediction, which are highly related to revenue.
An essential task proposed by us for evaluating these industrial capabilities is action prediction, which includes \textbf{next-item} and \textbf{next-attribute} predictions, the pivotal focus of user sequence modeling.
Traditional next-item prediction involves inputting a user interaction sequence into the model and outputting the item the user will most likely interact with next time from the candidate items.
Two key differences exist between our proposed task and the traditional next-item prediction task.
First, in our approach, we use the compressed user sequence, represented by user embeddings, as a soft prompt for LLMs to generate predictions rather than directly using the original user sequence.
Additionally, we expand the prediction target beyond the item itself to include other attributes (i.e., side information) of the next items, such as title, category, brand, and shop. 
For example, we may ask, "\textit{what item brand is the user most likely to click at next time?}"
Such capability of providing more diverse responses is a significant advantage of LLMs.


\subsubsection{\textbf{Interest Perception.}}
One revolutionary improvement of LLM-based recommendations compared to traditional recommendations should be introducing remarkable diversity. Restricted by the training paradigm and CF framework, traditional recommendations will concentrate on popular items and users with frequent interactions. We hope that user embeddings can assist LLM-based approaches in recalling diverse user interest items to improve personalization and enhance user experience. 
To evaluate the effect of user embeddings in enhancing personalization, we design some unique question prompts to assess how well LLMs prompted by user embeddings know the user's interest. 
As the interaction sequence may be noisy and overlong, the user embedding should highlight the core and indispensable parts (e.g., long- and short-term interests) and reveal the user's interest trajectory. 
Regarding user \textbf{long- and short-term interests}, we build the prompts such as "\textit{the user wants to purchase products from a certain category/brand/store, based on user's history information, recommends three options}," and "\textit{what is the user's recent interest category/brand/store}." The ground truth of the former is related to the whole interaction sequence, while the latter consists of the recently interacted items.  
Considering the user's \textbf{interest trajectory}, we may ask, "\textit{has the user's interest recently changed, and what is the trajectory of interest change}." 

\subsection{Technical Details}
Besides introducing the main evaluation process, several technical details merit attention, which we discuss below. 

\subsubsection{\textbf{Data Preprocessing}} 
\label{sec:data}
First, for privacy protection and compliance regulations, we anonymize the user IDs to prevent locating users and filter out sensitive items. Furthermore, the original user interaction data is lengthy and noisy, and the number of items is massive, which is inconducive to encoder models' training efficiency and stability. To address these issues, we conduct necessary data preprocessing on both the user and item sides. 
On the user side, we filter out users with interaction sequences that are either too long or too short, thereby enhancing the efficiency and stability of mini-batch training.
Regarding items, we remove those with fewer than three user interactions to reduce the size of the item pool, significantly improving the training effectiveness of user encoders.

\subsubsection{\textbf{Prompt Designing}} 
\label{sec:prompt}
We design templates of question prompts to enhance evaluation efficiency and objectivity. Given a task-specific template and user data, we can automatically generate the corresponding question and answer based entirely on user interactions. 
For instance, a template designed for the direct feature task might be "\textit{what are the last $k$ items the user has clicked?}", and inputting the interaction sequence $s_i = [v_1,\cdots,v_{n_i}]$ of user $u_i$ would automatically yield the answer $v_{n_i-k+1},\cdots,v_{n_i}$. 
Since it is unrealistic for LLMs to generate complete answers to highly specialized questions, we provide candidates for such questions. 
Furthermore, we adopt some filter rules to avoid trivial or overly burdensome questions, for example, discarding questions with an answer of $0$ (i.e., no match features) in match feature tasks and ensuring $k$ is less than $4$ in direct feature tasks. 
Some necessary instructions are also provided to guide LLMs' generation, like arranging multiple answers in chronological order for direct feature tasks or sorting the answers by user interest level for interest perception tasks. However, these instructions only play an auxiliary role, as the fine-tuning stage plays the leading role.

\begin{table*}
\caption{\label{tab:overall}
    Results of evaluating LLMs on \name dataset. The overall average (Avg.) score is bold in each row.
}
\begin{threeparttable}
\centering
\begin{tabular}{
  l 
  l 
  >{\centering\arraybackslash}p{0.8cm} 
  >{\centering\arraybackslash}p{0.8cm} 
  >{\centering\arraybackslash}p{0.8cm} 
  >{\centering\arraybackslash}p{0.8cm} 
  >{\centering\arraybackslash}p{0.8cm} 
  >{\centering\arraybackslash}p{0.8cm} 
  >{\centering\arraybackslash}p{0.8cm} 
  >{\centering\arraybackslash}p{0.8cm} 
  >{\centering\arraybackslash}p{0.8cm} 
  >{\centering\arraybackslash}p{0.8cm} 
  c 
}
    \toprule
    \toprule
    \multirow{2}{*}{Methods} & \multirow{2}{*}{Models} &
    \multicolumn{3}{c}{Sequence understanding} & \multicolumn{3}{c}{Action prediction} & \multicolumn{4}{c}{Interest perception} & \multirow{2}{*}{Overall Avg.} \cr
    & & DF & MF & Avg. & IP & AP & Avg. & LI  & SI & IT & Avg. & \cr
    \cmidrule(lr){1-13}
    \multirow{2}{*}{Text-based}
    & Text20 
    & 38.25 & 28.91 & 37.52 & 44.81 & 40.58 & 42.45 & 63.82 & 79.96 & 62.79 & 71.10 & \textbf{53.02} \cr
    \cmidrule(lr){2-13}
    & Text50 
    & 44.22 & 32.03 & 43.27 & 45.58 & 38.04 & 41.39 & 79.27 & 89.80 & 79.63 & 84.29 & \textbf{59.32} \cr
    \cmidrule(lr){1-13}
    \multirow{6}{*}{Emb-based}
    & GRU4Rec 
    & 33.67 & 23.44 & 32.86 & 34.94 & 30.02 & 32.20 & 46.98 & 41.33 & 45.56 & 43.97 & \textbf{37.15} \cr 
    \cmidrule(lr){2-13}
    & SASRec 
    & 40.41 & 24.22 & 39.14 & 42.49 & 35.27 & 38.47 & 59.20 & 53.32 & 56.39 & 55.72 & \textbf{45.66} \cr
    \cmidrule(lr){2-13}
    & Mamba4Rec 
    & 30.54 & 18.75 & 29.62 & 34.25 & 28.10 & 30.83 & 38.82 & 34.69 & 35.64 & 36.06 & \textbf{32.69} \cr
    \cmidrule(lr){2-13}
    & HSTU 
    & 43.29 & 25.00 & 41.86 & 47.30 & 36.53 & 41.31 & 70.85 & 60.06 & 65.67 & 64.45 & \textbf{50.88} \cr
    \cmidrule(lr){2-13}
    & Trm++ 
    & 47.28 & 27.34 & 45.72 & 48.67 & 37.35 & 42.38 & 75.25 & 64.14 & 70.10 & 68.71 & \textbf{53.88} \cr
    \bottomrule
    \bottomrule
\end{tabular}
\end{threeparttable}
\end{table*}

\subsection{In-depth Analysis: Scaling Law}

We aim to reveal the scaling law of using embeddings to prompt LLMs for personalization, which benefits resource distribution in industrial scenarios. Following valuable experiences from LLMs and recommender systems, we focus on two factors: \textbf{model size} (of user encoders) and \textbf{sequence length} (of interaction data in the pre-training stage).
Scaling up the model size is validated to improve model performance in various domains, e.g., NLP~\cite{kaplan2020scaling}. 
In addition, the previous experience in SRs~\cite{fang2020deep} is that scaling up the sequence length can consistently improve the recalling and ranking performance. We will explore whether such conclusions still hold for the effect of user embeddings in prompting LLMs for personalization.  
As the core of LLMs, Transformers are proven to possess excellent scaling capabilities. Therefore, our scaling experiments will apply Transformers as the backbone models.

\section{Experiments}

We assess the performance of various user embeddings in prompting LLMs for personalized question answering. Specifically, we want to reveal the answers to the following research questions:
\begin{itemize}[leftmargin=*]
\item \textbf{RQ1}: What is the overall performance of user embeddings in prompting LLMs for personalized question answering?
\item \textbf{RQ2}: How is the contribution of each component?
\item \textbf{RQ3}: What is the scaling law of user embeddings in prompting LLMs for personalized question answering?
\item \textbf{RQ4}: How is the efficiency of embedding-based GRs?
\end{itemize}

\subsection{Experimental Setup}
\subsubsection{\textbf{Datasets}}
We randomly split $31,317,087$ interactions from source data into $9:1$ for the training and test sets. The training set serves in the pre-training and fine-tuning stages. Based on the test set, we generate $7,192$ personalized Q\&A  for the evaluating stage.
\subsubsection{\textbf{Models}}
The user embeddings to be evaluated are produced by encoder models. 
We experiment on several state-of-the-art models to comprehensively evaluate embedding-based GRs. 
\begin{itemize}[leftmargin=*]
\item \textbf{GRU4Rec}~\cite{hidasi2015session}: leverages the gated recurrent units (GRUs), the famous class in recurrent neural networks (RNNs) to learn the user sequential representation.
\item \textbf{SASRec}~\cite{kang2018self}: applies self-attention, the core of transformers, to embed user interests from interaction sequences.
\item \textbf{Mamba4Rec}~\cite{liu2024mamba4rec}: captures user temporal dynamics using state space models (SSMs). Employing a selective mechanism, it identifies important states while ignoring irrelevant ones. 
\item \textbf{HSTU}~\cite{zhai2024actions}: redefines recommendations as sequential transduction tasks within GRs. It introduces a novel module, HSTU, specifically designed for industrial recommendation scenarios.
\item \textbf{Trm++}~\cite{zhai2024actions}: is a prominent Transformer variant~\cite{touvron2023llama} leverages upgraded architecture accompanied by RoPE~\cite{su2024roformer}, SwiGLU.
\end{itemize}
As illustrated in Figure~\ref{fig:overview}, the user embeddings will serve as the soft prompts to contextualize LLMs to generate personalized responses conforming to user interests. To guarantee fairness, we utilize a uniform backbone LLM, Qwen2.5-3b-instruct~\cite{yang2024qwen2}, for all encoder models. We also consider text-based GRs for comparison as follows.
\begin{itemize}[leftmargin=*]
\item \textbf{Text20 \& Text50}: leverages user sequence of $20$ and $50$ interactions, respectively, as text prompt of LLMs to generate responses.
\end{itemize}

\subsubsection{\textbf{Implementation Details}}
Following the previous research~\cite{sun2019bert4rec,kang2018self}, we initialize the trainable parameters using Gaussian distribution. We employ the AdamW optimizer~\cite{loshchilov2017decoupled} with a learning rate of 1e-4 and mini-batch size $256$ to optimize all models. Based on the configurations in prior studies~\cite{sun2019bert4rec, zhang2019feature, kang2018self} and maintaining comparability, we ensure that all model sizes are on the same order of magnitude (approximately 20M) and globally adjust the hyperparameters. For instance, regarding GRU4Rec, we utilize a $3$-layer GRU module with a hidden size of $1024$, and for SASRec, the number of layers is set to $L=6$, each with $h=8$ attention heads and dimension size $d=1024$, with an inner size (i.e., of FNN layers) of $1536$. We utilize $N=512$ as the global sequence length. 
We zero-pad sequences shorter than $N$ and truncate those longer than $N$.
The input and output layers share an item embedding matrix, and the dropout rate is set to $0.1$ to alleviate overfitting. The maximum numbers of epochs for pre-training and fine-tuning are $10$ and $3$, respectively. 
Our implementations\footnote{We release code and data at~\url{https://github.com/OpenStellarTeam/UQABench} to encourage reproducibility and facilitate further research on this topic.} are conducted in the environment using PyTorch 2.4.0 and Python 3.10.13 with $8\times$ GPUs.

\subsection{Overall Comparison (RQ1)}

We evaluate several advanced user encoder models on proposed \name following the standardized evaluation process (Figure~\ref{fig:overview}) and report the overall performances in Table~\ref{tab:overall}. In general, \name shows excellent discriminative power for different models, demonstrating its evaluation effectiveness. We draw some interesting observations as follows:

\begin{itemize}[leftmargin=*]

\item \textbf{Embedding-based GRs \textit{vs.} Text-based GRs}: 
The best model of embedding-based GRs (i.e., using Trm++) can achieve satisfactory performance on par with text-based GRs, showing the effectiveness of this method in personalization. Remarkably, embedding-based GRs surpass text-based counterparts in item prediction tasks since they are pre-trained to capture more prior information. However, embedding-based GRs perform significantly worse than text-based methods in interest perception tasks, indicating room for improvement in the framework.

\item Among embedding-based methods, HSTU and Trm++ present superior performance, providing high-quality user embeddings to prompt LLMs to give relatively satisfied responses, conforming to user preferences.  
The model performances exhibit anisotropy, meaning the performance of models on each task is not linearly correlated. For example, SASRec performs close to HSTU on sequence understanding and action prediction tasks yet far worse than HSTU on interest perception.


\item Regarding evaluation tasks, most encoder models perform significantly well in interest perception tasks, showcasing that the LLM-based system is inclusive and diverse to perceive user personalized interest instead of only popular items. The models' performance on action prediction tasks is also quite acceptable, suggesting that the embedding-based GRs paradigm has potential in traditional recommendation tasks. Nevertheless, the sequence understanding task results are not good enough, especially the sub-task match feature, reflecting that the current framework is immature to completely substitute the industrial feature engineering and needs future improvements.



\end{itemize}

\subsection{Ablation Study (RQ2)}

\begin{table}[t]
    \caption{Ablation in pre-training stage.}
        \vspace{-2mm} 
    \begin{threeparttable}
    \begin{tabular}{cccccc}
        \toprule
        \toprule
        \multirow{1}{*}{Training}&\multirow{1}{*}{Inputs}&
        SU & AP & IP & Avg. \cr
        \cmidrule(lr){1-6}
        \multirow{3}{*}{SL}
        & Full Info             & 45.72 & 42.38 & 68.71 & 53.88 \cr
        & \textit{w/o} Text     & 42.08 & 39.27 & 66.84 & 51.16 \cr
        & \textit{w/o} Side     & 39.55 & 37.83 & 61.47 & 47.87 \cr
        \cmidrule(lr){1-6}
        \multirow{3}{*}{CL}
        & Full Info             & 42.41 & 39.25 & 62.37 & 49.40 \cr
        & \textit{w/o} Text     & 37.06 & 35.35 & 58.02 & 44.99 \cr
        & \textit{w/o} Side     & 36.50 & 32.84 & 51.96 & 41.47 \cr

        \bottomrule
        \bottomrule
    \end{tabular}
    \end{threeparttable}
        \label{tab:ablation_1}
        \vspace{-2mm}
\end{table}

\begin{table}[t]
    \caption{Ablation in fine-tuning stage.}
        \vspace{-2mm} 
    \begin{threeparttable}
    \begin{tabular}{cccccc}
        \toprule
        \toprule
        \multirow{1}{*}{Integration}&\multirow{1}{*}{Strategy}&
        SU & AP & IP & Avg. \cr
        \cmidrule(lr){1-6}
        \multirow{2}{*}{Mean Pooling}
        & Full FT               & 43.70 & 38.83 & 72.70 & 53.75 \cr
        & Adapter      & 45.72 & 42.38 & 68.71 & 53.88 \cr
        \cmidrule(lr){1-6}
        \multirow{2}{*}{Q-Former}
        & Full FT               & 45.35 & 44.37 & 69.89 & 55.00 \cr
        & Adapter      & 31.15 & 29.92 & 31.80 & 30.96 \cr
        \bottomrule
        \bottomrule
    \end{tabular}
    \end{threeparttable}
        \label{tab:ablation_2}
        \vspace{-2mm}
\end{table}

To guarantee evaluation reliability and fairness, we conduct an ablation study in terms of pre-training and fine-tuning stages to determine the optimal settings of the experiments.
\subsubsection{\textbf{Pre-training}}
\label{sec:ablation_1}
The two critical factors in pre-training are model inputs and training methods. We consider three variants of model inputs: (1) \textbf{Full Info}: inputs entire information, including item ID and side information (e.g., category ID and category name), (2) \textbf{\textit{w/o} Text}: leverages the item ID and ID-based side information (e.g., category ID), meaning without textual information, and (3) \textbf{\textit{w/o} Side}: only uses item ID without side information. Furthermore, we consider two training methods: (1) \textbf{SL}: uses the CE loss to supervise training, and (2) \textbf{CL}: utilizes InfoNCE loss to perform contrastive training on encoders. According to the results in Table~\ref{tab:ablation_1}, we observe that the combination of \textbf{Full Info} + \textbf{SL} yields the best performance. Consequently, we utilize such a setting for our experiments.


\subsubsection{\textbf{Fine-tuning}}
\label{sec:ablation_2}
We explore the optimal settings in fine-tuning.
We compare two compression approaches in the adapter: (1) \textbf{Mean Pooling}: uses a mean pooling layer to compress user sequence embeddings into a single embedding, (2) \textbf{Q-Former}: leverages Q-Former to generate fixed-length embeddings, set to 16 by default. 
Moreover, we consider two fine-tuning strategies: (1) \textbf{Full FT}: fine-tunes the adapter and LLM together, and (2) \textbf{Adapter}: fine-tunes only the adapter. Table~\ref{tab:ablation_2} shows that the setting of \textbf{Full FT} + \textbf{Q-Former} achieves optimal performance. However, we observe that the Q-Former method is sensitive to the hyperparameters (e.g., learning rate), leading to instability (i.e., emerging NaN loss). Therefore, we opt for a more stable configuration, \textbf{Mean Pooling} + \textbf{Adapter}, as our experiment's essential fine-tuning configuration for consistency and reliability.

\begin{table}[t]
    \caption{The scalability in terms of model size.}
        \vspace{-2mm} 
    \begin{threeparttable}
    \begin{tabular}{cccccc}
        \toprule
        \toprule
        \multirow{2}{*}{Model size}&\multirow{1}{*}{Architecture}&
        \multicolumn{4}{c}{Evaluation tasks} \cr
        \cmidrule(lr){3-6}
        & $(l, d, h)$ & SU & AP & IP & Avg. \cr
        \cmidrule(lr){1-6}
            3.2M   & (4, 256, 4)     & 39.00 & 34.74 & 58.16 & 45.26\cr
            13M    & (4, 512, 8)     & 38.62 & 37.08 & 67.55 & 49.86\cr
            25M    & (8, 512, 8)     & 45.72 & 42.38 & 68.71 & 53.88\cr
            57M    & (8, 768, 12)    & 45.29 & 42.15 & 74.74 & 56.16\cr
            85M    & (12, 768, 12)   & 47.12 & 44.32 & 76.48 & 58.08\cr
            303M   & (24, 1024, 8)   & 50.43 & 46.65 & 77.30 & 60.01\cr
            680M   & (24, 1536, 12)  & 52.20 & 48.67 & 83.74 & 63.78\cr
            1.2B   & (24, 2048, 16)  & 53.98 & 49.16 & 84.97 & 64.86\cr
     
        \bottomrule
        \bottomrule
    \end{tabular}
    \end{threeparttable}
        \label{tab:scale_size}
        \vspace{-2mm}
\end{table}


\begin{figure}[t]
    \centering
    \vspace{-2mm}
    \includegraphics[width=0.75\linewidth]{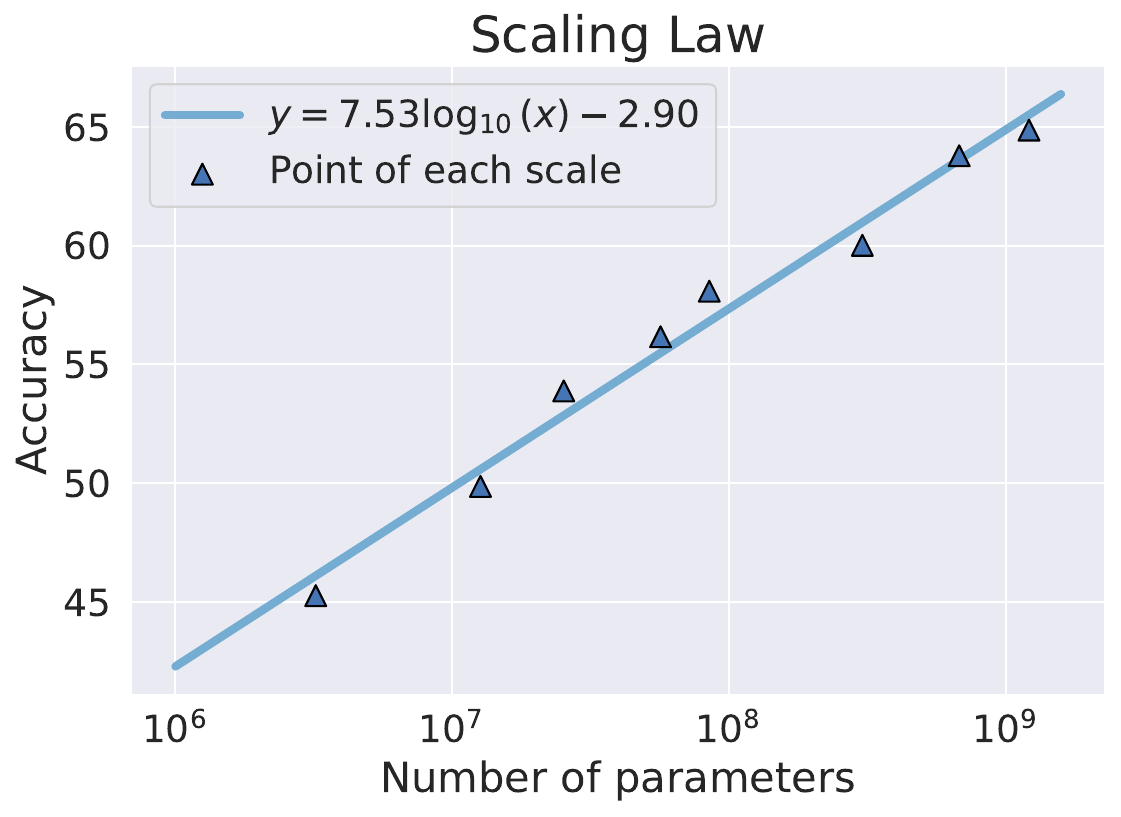}
    \vspace{-3mm}
    \caption{
    Scaling law in terms of the model size.
    }
    \label{fig:scaling_law}
    \vspace{-3mm}
\end{figure}

\subsection{Scalability Study (RQ3)}
We explore the scalability of user embeddings in prompting LLMs for personalization. We employ Trm++ as the backbone encoder and follow the proposed evaluation process. Our scalability study considers two critical factors of encoders: model size and sequence length. The experimental details and results are as follows.

\subsubsection{\textbf{Model Size}} 
We scale the model size of the encoders, ranging from 3M to 1.2B. The results are reported in Table~\ref{tab:scale_size}, where we specify the architectural details, i.e., the number of layers $l$, dimension size $d$, and the number of heads $h$, of the encoder. The experimental results show a clear scaling law between model performance and model size, illustrated in Figure~\ref{fig:scaling_law}. Such a result is significant to the application scenario: We can strengthen encoder models offline (i.e., scaling up model size) to consistently enhance the personalization performance of LLMs in the online environment without impairing the inference efficiency.

\subsubsection{\textbf{Sequence Length}} 
We also investigate the impact of sequence length on personalized performance by adjusting this parameter during pre-training. Specifically, we experiment with various maximum sequence lengths ranging from $32$ to $512$ and present the corresponding outcomes in Table~\ref{tab:scale_length}. We find that while embedding-based GRs enhance performance as the sequence length increases, this improvement rate diminishes beyond a certain length (e.g., $256$). This trend is reasonable, given that the actual length of user interactions is generally limited.


\begin{table}[t]
    \caption{The scalability in terms of sequence length.}
        \vspace{-2mm} 
    \begin{threeparttable}
    \begin{tabular}{ccccc}
        \toprule
        \toprule
        \multirow{2}{*}{Sequence length}&
        \multicolumn{4}{c}{Evaluation tasks} \cr
        \cmidrule(lr){2-5}
        & SU & AP & IP & Avg. \cr
        \cmidrule(lr){1-5}
            32    & 27.78 & 30.83 & 38.79 & 33.38 \cr
            64    & 37.21 & 35.67 & 53.75 & 43.40 \cr
            128   & 40.88 & 38.64 & 63.97 & 49.49 \cr
            256   & 42.41 & 41.23 & 69.33 & 52.96 \cr
            512   & 45.72 & 42.38 & 68.71 & 53.88 \cr
        \bottomrule
        \bottomrule
    \end{tabular}
    \end{threeparttable}
        \label{tab:scale_length}
        \vspace{-2mm}
\end{table}

\subsection{Efficiency Study (RQ4)}

We compare the efficiency of text-based and embedding-based GRs to provide insights into deployment scenarios. When deploying LLMs for online serving, the number of tokens is critical to inference efficiency. Therefore, we report the average number of tokens for the methods tested, as shown in Table~\ref{tab:efficiency}, where \textbf{Trm-M} and \textbf{Trm-Q} correspond to Trm++ leveraging two compression methods (\textbf{Mean Pooling} and \textbf{Q-former}). The results indicate that embedding-based GRs significantly reduce the number of tokens compared to their text-based counterparts, achieving increased efficiency by a factor of \(8\times\) to \(19\times\). This efficiency gain is achieved by compressing the user context, i.e., interactions, into fewer tokens (1 for Trm-M and 16 for Trm-Q). In the future, refining question prompts may substantially increase efficiency as they occupy relatively more tokens than soft prompts (i.e., user embeddings).


\begin{table}[t]
    \caption{Efficiency comparison on UQABench.}
        \vspace{-2mm} 
    \begin{threeparttable}
    \begin{tabular}{cccc}
        \toprule
        \toprule
        \multirow{1}{*}{Methods}&\multirow{1}{*}{Models}&
        Performance & Avg. Tokens \cr
        \cmidrule(lr){1-4}
        \multirow{2}{*}{Text-based}
        & Text20               & 53.02 & 1171.23  \cr
        & Text50               & 59.32 & 2498.19  \cr
        \cmidrule(lr){1-4}
        \multirow{2}{*}{Emb-based}
        & Trm-M                & 53.88 & 133.28  \cr
        & Trm-Q                & 55.00 & 148.28  \cr
        \bottomrule
        \bottomrule
    \end{tabular}
    \end{threeparttable}
        \label{tab:efficiency}
        \vspace{-2mm}
\end{table}

\section{Related Work}
\subsection{Sequential Recommendations}
Sequential recommendations (SRs) learn the user representation from the historical interaction sequence, then calculate the recommendation scores of candidate items and choose top-$k$ as the recommendation result~\cite{fang2020deep,wang2019sequential}. 
Previous works try various deep learning modules to enhance user modeling performance. For example, GRU4Rec~\cite{hidasi2015session} leverages the GRU layers, Caser~\cite{tang2018personalized} uses the CNN layers, and HGN~\cite{ma2019hierarchical} adopts GLU layers. The recent research direction of SRs gradually converges to prevalent Transformer~\cite{vaswani2017attention}. The Transformer-based methods, such as SASRec~\cite{kang2018self}, BERT4Rec~\cite{sun2019bert4rec}, FDSA~\cite{zhang2019feature}, BST~\cite{chen2019behavior}, and CORE~\cite{hou2022core}, leverage attention layers to capture the node correlation within the interaction sequence, which remarkably elevates the representation learning.
Despite the performance effectiveness of Transformer-based methods, the main challenges are efficiency~\cite{tay2022efficient}. The attention module of Transformers brings quadratic computational complexity with sequence length, which is unrealistic for deployment in industrial recommender systems that need low inference time. The most recent works, such as LinRec~\cite{liu2023linrec}, LRURec~\cite{yue2024linear}, Mamba4Rec~\cite{liu2024mamba4rec}, provide efficient solutions that can par with or even outperform Transformer-based methods. However, one essential issue is that the CF framework of SRs may lead to information cocoons and jeopardize user experience.

\subsection{Generative Recommendations}
Generative recommendations (GRs) leverage the generative capabilities of LLMs to generate more personalized and diverse results besides candidate items~\cite{xu2024prompting,zhao2023recommender,zhang2023recommendation,wu2024survey}. The mainstream of GRs, text-based GRs, leverage the text information in user interaction sequences as the user context to prompt LLMs to generate personalized results ~\cite{liu2023chatgpt,petrov2023generative,kang2023llms,geng2022recommendation,lyu2023llm}. 
Some frequently used prompting methods in the LLM area can be adapted to text-based GRs, such as in-context learning (ICL)~\cite{gao2020making}, and chain-of-thought (CoT)~\cite{wei2022chain}. The ICL-based methods~\cite{liu2023chatgpt,li2023bookgpt} incorporate the task description and in-context demonstrations (i.e., few-shot). While the CoT-based approaches~\cite{huang2023recommender,wang2023recmind} design task-specific reasoning steps to assist LLMs in generating personalized responses step-by-step. 
Nevertheless, employing the whole user interaction sequence as the text context of LLMs is impractical and unscalable. A regular user interaction sequence with hundreds of items can produce 10k to 100k tokens, which may cause an unacceptably high inference time and cost and even exceed the LLMs' context window limit. Therefore, a line of GRs tries to transform the user interaction sequence into a compact form, user embedding, to softly prompt LLMs for personalization~\cite{li2023prompt,ning2024user,li2023personalized,hebert2024persoma}. The embedding-based GRs highly reduce the token number compared to directly using text information, improving the efficiency and scalability of GRs. 



\section{Conclusion}
In the LLM era, the user interaction data provide a rich source to contextualize LLM to be more personalized, which implies a promising recommendation paradigm, GRs. By compressing the interaction sequence into a more compact form, user embeddings, the embedding-based GRs outperform text-based GRs with less inference pressure and are friendly to real-world deployments.  
To address the concerns about the effectiveness of embedding-based GRs, we propose a novel benchmark, \name, evaluating user embeddings from the personalization perspective. We propose a standardized three-stage evaluation process, including pre-training, fine-tuning, and evaluating. Furthermore, to comprehensively assess the quality of user embedding, we design three evaluation tasks and corresponding sub-tasks. We extensively experiment on state-of-the-art encoder models to explore the embedding-based GRs' effectiveness and provide some insights.


\newpage
\bibliographystyle{ACM-Reference-Format}
\bibliography{custom}

\end{document}